# Anisotropy in Antiferromagnets

K O'Grady[1], J Sinclair[1], K Elphick[1], R Carpenter[1,3], G Vallejo-Fernandez,[1] M. I. J. Probert,[1] and A Hirohata[2]

[1]Department of Physics, University of York, Heslington, York, YO10 5DD, UK
[2]Department of Electronic Engineering, University of York, Heslington, York, YO10 5DD, UK
[3]now at IMEC, Kapeldreef 75, 3001,Leuven, Belgium

**Abstract**

Due to the advent of antiferromagnetic (AF) spintronics there is a burgeoning interest in AF materials for a wide range of potential and actual applications. Generally, AFs are characterized via the ordering at the Néel temperature ($T_N$) but, to have a stable AF configuration, it is necessary that the material have a sufficient level of anisotropy so as to maintain the orientation of the given magnetic state fixed in one direction. Unlike the case for ferromagnets there is little established data on the anisotropy of AFs and in particular its origins and those factors which control it. In this paper these factors are reviewed in the light of recent and established experimental data. Additionally, there is no recognized technique for the first principle's determination of the anisotropy of an AF which can only be found indirectly via the exchange bias phenomenon. This technique is reviewed and in particular the implications for the nature of the anisotropy that is measured and its distribution. Finally, a strategy is proposed that would allow for the development of AF materials with controlled anisotropy for future applications.

1.    **Introduction**

The phenomenon of antiferromagnetism was first formally described by Louis Néel in the early 1930s who suggested that in a material with atoms having a finite and often quite large moment, it would be possible for the moments on individual atoms to align antiparallel in a similar manner to the parallel alignment that occurs in ferromagnets [1,2]. Néel also suggested that such magnetic order could be disrupted by thermal energy and defined the ordering parameter (θ) to occur at a fictitious negative temperature in a pseudo Curie-Weiss law where χ is the susceptibility, C is a constant and T is temperature. In addition to this mathematical definition of the phenomenon shown in equation 1, Néel also defined a finite positive temperature, $T_N$, at which the ordering will be observed in practice.

$$\chi = \frac{C}{T+\theta} \qquad (1)$$

This temperature is now known as the Néel temperature ($T_N$) and all these relevant temperatures and the variation of inverse of susceptibility with temperature are shown in Figure 1 which is the classical behavior of an antiferromagnetic material.

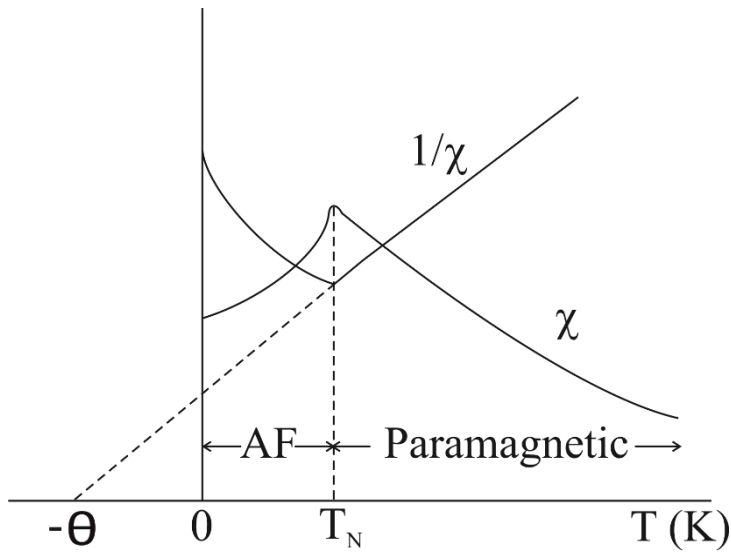

Figure 1. Temperature variation of the AF susceptibility.

In practice the cancellation of the moments very often occurs along crystallographic planes such that within a given plane the ordering is ferromagnetic (F) in nature but then the spins on neighboring planes order in the opposite direction thereby giving complete cancellation of the moment. Such materials are known as sheet AFs. However under certain circumstances the moment cancellation is imperfect leading to ferrimagnetic order which occurs mainly in mixed metal oxides. The two different forms of AF order are shown in Figures 2(a) and (b).

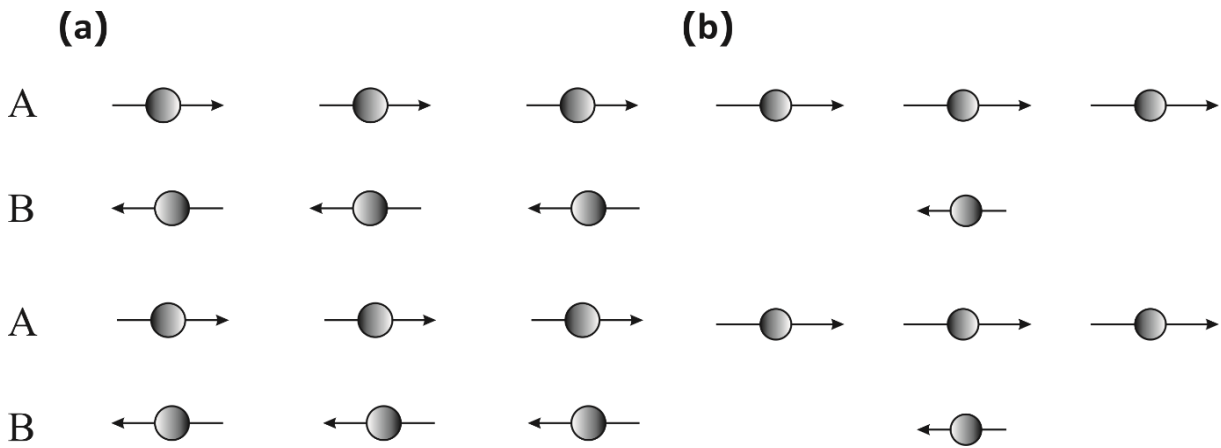

Figure 2. Magnetic moment arrangement along two crystallographic planes, A and B, for (a) an antiferromagnetic and (b) a ferrimagnet.

Under such circumstances a cancellation of the atomic moments occurs such that the material appears to be non-magnetic. However, at normal temperatures and in very large fields it is possible that some slight misalignment or canting of the spins on the neighboring sub-lattices can occur giving rise to a finite susceptibility as shown in Figure 1.

The subject of antiferromagnetism was of only academic curiosity until the discovery of the exchange bias phenomena by Meiklejohn and Bean in 1956 [3]. Exchange bias occurs when an AF material is grown in contact with an F material such that the coupling between the spins at the surface of the AF material and the F material align across the interface thereby pinning the spin in the surface layers of the F material. This then induces an additional unidirectional anisotropy into the F layer. Under these circumstances a shifted hysteresis loop results when the system is cooled from above the Néel temperature of the AF to a temperature below $T_N$. Also, a significant enhancement to the coercivity of the loop results.

Figure 3(a) shows the original result of Meiklejohn and Bean which was obtained for a system of partially oxidized cobalt particles in a bath of mercury which was frozen in a field to 77K. In this case CoO is the AF and Co is the F, giving rise to the measured magnetic moment. Similar effects can be observed particularly in thin film structures where an AF layer is grown either beneath or immediately above an F layer. Correct selection of materials can result in a hysteresis loop where both values of the pseudo-coercivity ($H_c$) can lie in (say) a negative field following the system being field cooled in a positive field. Such a structure is shown in Figure 3(b) with the resulting magnetization curve in Figure 3(c) [4]. This type of system is used to pin one F layer in the tunnelling magnetoresistive (TMR) devices that are used in all read heads in hard drives at this time [5].

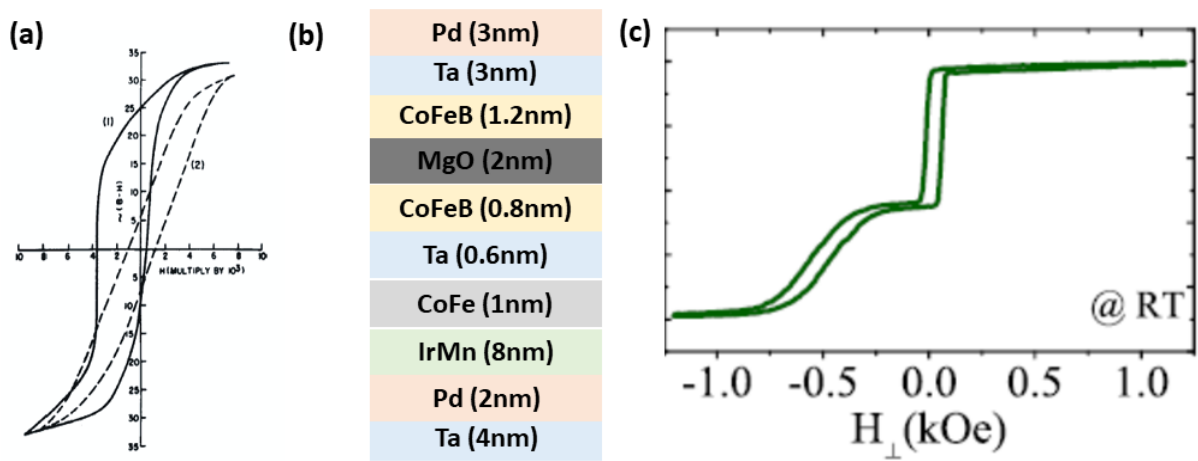

Figure 3. (a) Original data obtained by Meiklejohn and Bean on CoO/Co nanoparticles, (b) structure of a typical TMR device and (c) Typical hysteresis loop for a structure similar to those found in read head TMR devices [4].

Importantly the recent discovery of a spin Hall effect in AF materials [6] has led to the rapid development of a new field of endeavor whereby spintronic devices can now be designed and fabricated using AF materials [7]. Hence the entire field of the study of AF materials is currently the most active in the field of magnetism.

The concept of spin ordering in an AF material be it via atom to atom antiparallel alignment or a sheet structure implies that in some way there must be either an easy direction or an easy plane in the materials in a similar manner to that which occurs in ferromagnets. This in turn implies that in some way AF materials exhibit anisotropic behavior which must be solely based on the anisotropy of the exchange interaction as there can be no shape or other effects due to the absence of a net magnetic moment. Hence the anisotropy is magneto-crystalline in origin. This implies that the direction of the ordering can be altered if, in some way, a potential can be applied which will couple to the spins. Of course such a potential cannot be a conventional applied magnetic field and one way in which a reordering of the spin orientation can occur is via an exchange couple where an F layer is grown adjacent to the AF layer in a thin film. Changing the orientation of the F layer and field annealing then generates an "exchange field" that can orient the direction of the AF. This occurs when the annealing temperature $T_{set}>T_N$ or at a lower temperature by thermal activation of the orientation of the AF [5]. This is the basis of the exchange bias effect.

## 2. Reversal in Antiferromagnets

Given that it is possible to vary the orientation of AF alignment this automatically leads to the concept of the reversal mechanism that is present in such a system. Again it was Louis Néel who postulated the existence of a domain structure in an AF material [8]. The origin of such a domain structure is not clear because in an F the existence of domains is driven by the magnetostatic energy when a material is magnetized and no such potential would exist in an AF material. However were the AF material part of an exchange bias system, then any domain structure in the F layer could print through to the AF layer via the exchange coupling. Indeed the existence of AF domains has been verified by both optical studies [9] and more recently using X-ray techniques [10].

Of course and, particularly for the case of technological applications, the concept of an AF domain leads to the concept of a critical size at which the material would exist as a single domain grain. This is particularly applicable for granular thin films grown by sputtering or other ion beam techniques and is of importance because it is these films that are used computer hard disk in read heads and potentially other AF spintronic devices. The critical size at which an AF material would transform from a multi domain state to a single domain state is not known and again the absence of a magnetostatic energy makes it unclear as to the nature of the driving force that would give rise to such a transition. The naïve approach might be to say that the transition would occur in a grain of dimensions such that a domain wall could not be accommodated. Hence there are two distinct cases that must be considered.

### 2.1 The Multi Domain Case

Since about the year 2000 there have been a number of works based on numerical modelling in which the structure and behavior of domains and domain walls in AF materials have been investigated. In the space available it is not possible to describe all the models but some of those most prominent and most pertinent to work that follows include the work of Stiles and McMichael [11] who produced a model with exchange coupled F grains and exchange decoupled grains in the AF layer. In this sense the model was the closest to the actual structures used e.g. in read heads. The model also included thermal activation and was based on a critical angle between the alignment of the AF layer and the F layer at which thermal energy would induce a change in the orientation of the AF.

Stamps [12] suggested the existence of two mechanisms for exchange bias the first of which was due to a reversible formation of domain walls in the AF and the second due to irreversible processes leading to asymmetric hysteresis loops. The key result was the existence of higher order coupling terms when more than one AF sub lattice is present at the interface.

The dominant model in the early 2000s was that due to Nowak and co-workers [13] who proposed a domain state model in which the AF was a single crystal which generated domain walls that passed through impurities or defects in the lattice. A very large value of the anisotropy was assumed such that the AF domain wall width was very narrow, typically of atomic dimensions [13]. The domain walls become frozen-in during the setting process. The model was successful to the extent that it could predict the trend in the variation of the exchange bias ($H_{ex}$) with the film thickness but given that it was based upon a single crystal system it could not predict the lateral grain size dependence of the loop shift.

Once the York Model of Exchange Bias in polycrystalline thin films was published and became accepted [5] it became clear that it was also possible to explain features of exchange bias in single crystal or large grain polycrystalline samples using a simple domain wall pinning model. This model assumes that in multi-domain AF layers, defects and impurities will create domain wall pins which can be relatively strong due to the absence of magnetostatic energy which would displace the domain walls. Hence the strong domain wall pins have the effect of breaking up a single crystal or large grain film into progressively smaller grains such that eventually the smaller entities become magnetically a single AF domain and then reverse via rotation over an energy barrier rather than by the relatively easy process of domain wall motion. As the level of defects increases the resulting magnetic grain size can become small such that the orientation of the AF alignment becomes thermally unstable at finite temperatures of measurement. Hence the model predicts that as the transition occurs from multi-domain to single domain the loop shift will go through a peak declining as the material becomes thermally unstable [14]. Figure 4a shows data for a single crystal sample in which defects have been created by ion irradiation. Figure 4(b) shows calculated data based on the strong pinning model.

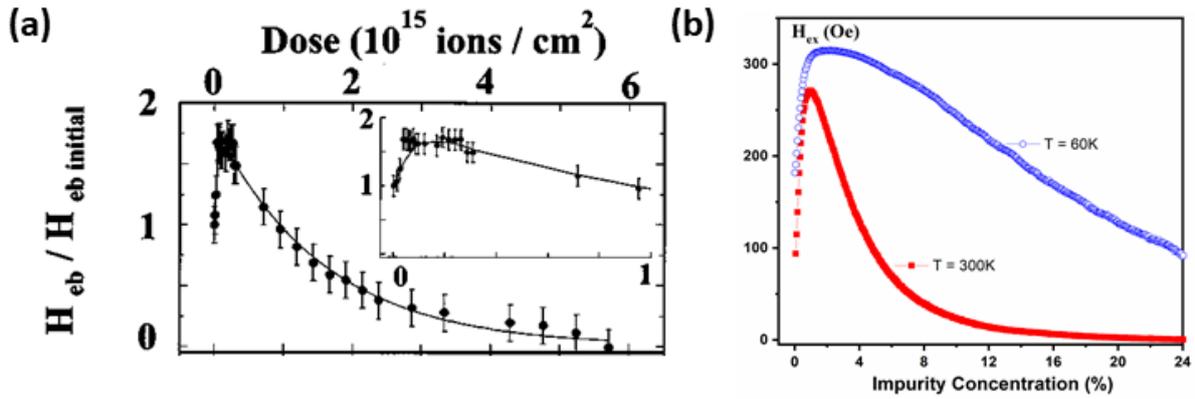

Figure 4. (a) Normalised exchange bias as a function of ion irradiation dose [15] at room temperature (inset is an enlarged view of the low dose region) and (b) Calculated exchange bias as a function of defect concentration using a strong pinning model at 300 and 60K, respectively [14].

Furthermore it is single domain grains that form the basis of all technological applications of AF materials currently in use and likely to be developed in the future. Hence an understanding of such systems is essential.

## 2.2 Single Domain Grains

The York Model of Exchange Bias is based upon the pioneering work of Stoner and Wohlfarth [16] in defining the behavior of single domain ferromagnetic grains. It incorporates the thermal activation theories of Néel [17], Street and Wooley [18] and Gaunt [19] for F grains and Fulcomer and Charap [20] who were the first to quantify thermal activation in AF grains.

As stated previously the exact critical size for single domain behavior in an AF material is not known but it seems likely to lie somewhere in the range from 30nm to 100nm and probably nearer the lower limit. Whilst there is no reason for incoherent reversal to occur as is the case in ferro and ferrimagnets, such behavior cannot be precluded. However almost all AF thin films deposited by conventional magnetron, RF or other form of ion beam sputtering, have sizes that lie in the range 5 to 20nm and it seems likely that all such grains will contain a single AF domain.

Following Stoner-Wohlfarth, the concept that the particles behave as single domains and behave in a uniaxial manner leads immediately to an energy barrier ΔE defined by

$$\Delta E = K_{AF} V (1 - H^*/H_K^*)^2 \qquad (2)$$

where $K_{AF}$ is the anisotropy constant of the AF grains and V their volume. $H^*$ is the exchange field from an F layer which is grown adjacent to the AF layer in an exchange bias system. $H_K^*$ is a pseudo anisotropy field arising from the presence of an easy direction in the AF grains and representing the energy barrier to reversal at T=0K.

Again, in a similar manner to ferromagnetic fine particles, the orientation of the anisotropy direction (not the AF order) is subject to thermal activation following a Néel Arrhenius type law of the form

$$\tau^{-1} = f_0\, exp - (KV/kT) \qquad (3)$$

where $f_0$ is an attempt frequency which we have previously reported to have a value of $2\times10^{12} s^{-1}$ for IrMn [21]. This leads to similar time dependence effects in the AF grains which, due to the width of the energy barrier distribution, generally follow a ln(t) variation [22]. To achieve exchange bias, i.e. a shifted hysteresis loop, it is first necessary to align or "set" the AF material. In principle this should be done above the Néel temperature, however in practice temperatures $T<T_N$ are used to avoid damage to the multilayer structure. Setting then proceeds by a thermal activation process following a ln(t) law. This can lead to incomplete setting of the AF. Similarly at normal temperatures unless great care is taken, smaller grains in the distribution may be thermally unstable and hence may also not contribute to the final value of the loop shift ($H_{ex}$). The resulting shift in the loop is then dependent upon the integral across the grain size distribution between the two limits set by the incomplete setting process and thermal disorder of smaller grains. This latter factor can be completely mitigated by undertaking all measurements at a temperature ($T_{NA}$) such that no thermal activation occurs [5].

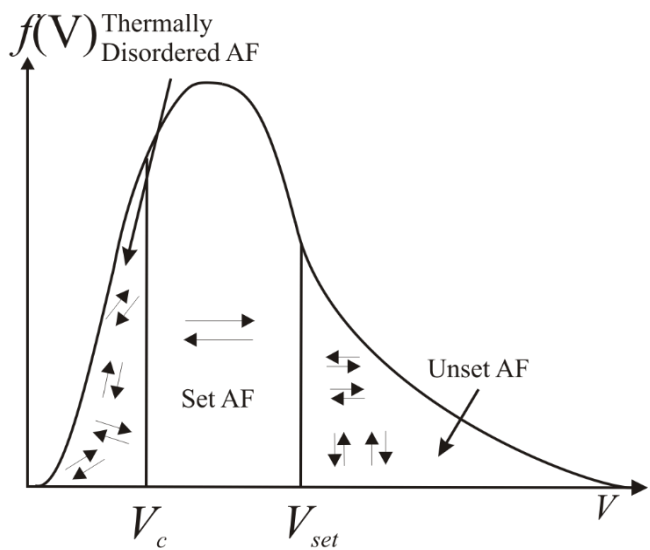

Figure 5. Schematic of the grain volume distribution in the AF layer after setting at a temperature $T_{set}$ in a magnetic field and cooling to a temperature $T_{meas}$ where a fraction of the AF is thermally unstable.

This situation is shown in Figure 5 where schematically, a measurement is made at a temperature somewhat above $T_{NA}$ and where the setting of the AF grains by a thermal activation process below $T_N$ has not induced complete alignment of the grains in the system. Under these circumstances the value of the loop shift $H_{ex}$ is given by

$$H_{ex} = C^* \int_{V_c}^{V_{set}} f(V)\, dV \quad (4)$$

where C* is a constant representing the stiffness of the AF to F coupling.

Controlled thermal activation of the smaller grains in the distribution with the magnetization of the F layer oriented in the opposite sense to that used for setting, now enables the determination of the effective anisotropy constant of the AF ($K_{AF}$). Following such a procedure with heating to progressively higher temperatures comes a point where the value of exchange bias becomes zero. At that point the grain size being thermally activated is equal to that of a grain having the median volume in the system which can be determined using transmission electron microscopy (TEM). The temperature at which $H_{ex}=0$ is then the median blocking temperature $<T_B>$. A value of $K_{AF}$ can then be found from

$$K_{AF} = \frac{ln(tf_0)k\,T}{<V>} \quad (5)$$

Here $<V>$ denotes the median volume of the grains in the film which can be determined by measurements of the grain diameters using a TEM and t is the time for which the applied field was held in an orientation opposite to that in which the sample was originally set. Note that for each measurement made, following the thermal activation time, the sample was re-cooled to a temperature $T_{NA}$ to ensure that no further reversal of the AF occurs during the measurement of the hysteresis loop.

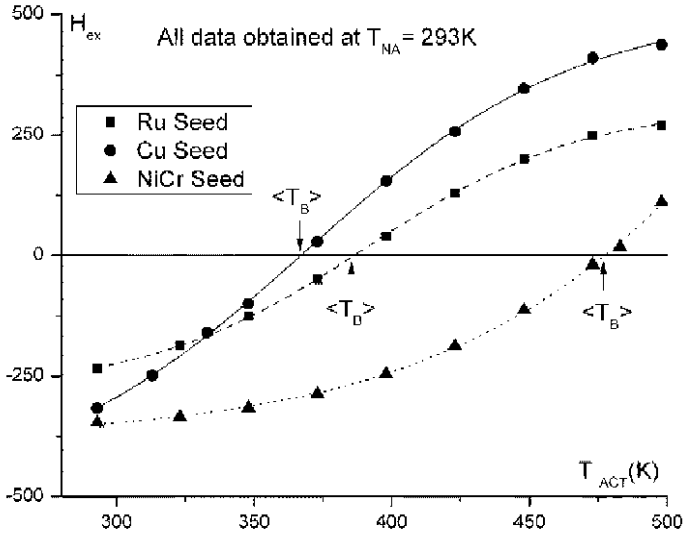

Figure 6. Example of a measurement of the distribution of blocking temperature curve for exchange biased system grown of different seed layers [23].

We have reported previously that the values of $K_{AF}$ obtained depends strongly on the crystallographic texture of the AF layer determined by the use of seed layers such that the (111) planes of the fcc IrMn lie in the plane of the substrate. For ideal texture such that no (111) planes can be detected out-of-plane, a value of $K_{AF}$ has been measured as large as $2.9 \times 10^7$ ergs/cc [23].

The solid lines in Figure 6 have been calculated directly from TEM measurements of the grain size distribution. In obtaining the fits by integration across the distribution, a constant value of $K_{AF}$ determined as described above has been used. This implies that there is no variation in the value of $K_{AF}$ from grain to grain otherwise the quality of resulting fits would not have been obtained. The anisotropy of IrMn and all other AFs is magnetocrystalline in origin and therefore, assuming that the grains are well crystallized, a constant value of $K_{AF}$ would be expected. However despite the fact that the system for which data is shown in Figure 6 grown on a NiCr seed layer is known to give good (111) texture in the plane of the film, the easy directions would be randomly oriented within the plane and therefore some distribution of the effective $K_{AF}$ would be expected. Similar fitting results have been obtained for many systems based on IrMn in our laboratories. Even when the degree of in plane texture was not as strong as was the case for this sample, a constant value for $K_{AF}$ gave a good fit. This apparent constant value of $K_{AF}$ is an unexplained feature of polycrystalline AF systems which needs to be explained.

## 2.3 Anisotropy in Antiferromagnets

As briefly discussed previously, AF behavior is commonly observed in transition metal oxides, certain metallic alloys generally containing Mn and also some of the rare earth metals at low temperatures. In the case of the transition metal oxides they almost all have an fcc structure with spins aligned parallel within the (100) planes but with each (100) plane oriented with the spins in an opposite sense to that in the neighboring (100) planes. Interestingly some of the intermetallic alloys such as IrMn, FeMn and MnNi have fcc structures and a similar spin structure associated with the (111) planes.

In the case of the transition metal oxide AFs, the nature of the ordering is derived from superexchange whereby the spin ordering in the transition metal occurs via the p orbital on the oxygen atoms. In all cases this leads to a relatively weak anisotropy. However the first AF material to be used to pin one layer in a GMR stack was NiO [24]. The thermal instability of this material, due to the low anisotropy, necessitated that every few hours a current was applied to the stack to both heat the NiO and generate a suitable magnetic field so that it could be reset. Hence after a relatively short space of time the use of FeMn became common as it had a significantly improved thermal stability due to it being more strongly anisotropic [25].

However FeMn exhibited significant corrosion issues and so from the mid 1990s onwards the alloy IrMn became the material of choice in all read head devices. In addition to its much improved corrosion resistance IrMn displayed remarkable thermal stability due to its very high anisotropy. An important and unexplained fact is that it is not the stoichiometric $IrMn_3$ alloy that is used as it induces a relatively modest exchange bias. Generally a composition much closer to $IrMn_4$ is found to give an optimum value for $H_{ex}$ [26]. This means that the structure must be at least partially disordered although it remains fcc. It has been reported and predicted that $IrMn_3$ may exhibit a triangular spin structure rather than exist as a planar AF material [27]. Such measurements have all been made on large single crystals or thick films and it is not clear, nor is it possible to determine, whether such a structure exists in thin film or indeed for the composition IrMn$_4$.

For all AFs it is the case that the anisotropy exhibited is magnetocrystalline in nature deriving from the spin orbit coupling in the lattice. It is the strength of the spin orbit coupling that directly leads to the magnitude of the anisotropy constant. However the nature of the magnetic anisotropy leads to its being temperature dependent and of the form [28]

$$K(T) = K(0)\left(1 - T/T_N\right) \qquad (6)$$

It is this temperature dependence that allows AFs in exchange bias systems to be set (aligned) at temperatures below T$_N$. The temperature dependence is also relatively rapid which also facilitates setting. However for all materials exhibiting cubic anisotropy the energy barrier to reversal ΔE=KV/4 or KV/12 [29] as compared with the uniaxial case for materials exhibiting a tetragonal structure where ΔE=KV. Hence it should be the case that materials exhibiting cubic anisotropy should be intrinsically thermally unstable in exchange bias systems. Hence it is curious that the material with the strongest known anisotropy is IrMn$_4$ which is cubic. A further curiosity occurs because in the York Model of Exchange Bias, excellent agreement is achieved when the underlying theory is based on the uniaxial case and the material studied was IrMn.

Of course in the ferromagnetic case there are other sources of anisotropy to be considered. These would include principally shape anisotropy, K$_s$, but also various stress related anisotropies. The occurrence of shape anisotropy would require the material to have a net moment as $K_s \alpha M_s^2$, where M$_s$ is the saturation magnetization of the material. Given that other than the possibility of uncompensated spins at the surface of a grain, the value of M$_s$ of an AF material is zero, and hence no significant shape anisotropy can exist. There is no known literature referring to the effects of stress on the anisotropy or other properties of an AF material but in general stress anisotropies are small and hence, whilst they may affect or influence the behavior of some of the transition metal oxides, they are unlikely to be significant compared to the value of the magnetocrystalline anisotropy in materials such as IrMn.

### 3. Experimental Measurements of Anisotropy

### 3.1 The Role of Texture

The main mechanism for controlling the anisotropy of AF layers lies in the control of the texture of the appropriate easy axis of the AF [23]. This is generally achieved by lattice matching between a seed layer having a cubic structure and for the case of IrMn, the (111) planes of the structure. In a previous work we examined three different seed layers being NiCr, Ru and Cu [23]. The original blocking temperature curves showing the value of <T$_B$> and hence the determination of K$_{AF}$ are shown in Figure 6.

| Seed Layer | $D_m$ (nm) | $\sigma$ (lnD) | $<T_B>$ (K) | $D_{set}$ (nm) | $K_{AF}$ (x$10^7$ ergs/cc) |
|---|---|---|---|---|---|
| NiCr | 3.9 | 0.42 | 477 | 4.1 | 3.3±0.4 |
| Ru | 6.0 | 0.38 | 386 | 7.6 | 0.94±0.06 |
| Cu | 10.7 | 0.37 | 367 | 14.0 | 0.28±0.02 |

Table 1. Summary of results for variation in $D_m$ (median grain size), standard deviation in lnD, $<T_B>$, largest grain size being set ($D_{set}$) and $K_{AF}$ at room temperature with seed layer.

In Table 1 are gathered the resulting values of the key parameters associated with these exchange bias systems. There is no direct correlation between the individual parameters $K_{AF}$ and $<D>$ because it is the product $K<V>$ that determines the value of $<T_B>$ and the growth rate of the different seed layers leads to a different grain size in the layer due to the columnar nature of the growth. Studies of the X-ray diffraction patterns of the three samples not reproduced here but contained in the original work, indicated that the IrMn grown on the NiCr seed layer had almost perfect in plane (111) texture whereas the sample grown on the Cu seed layer has almost perfect 3D random texture with the sample grown on Ru being an intermediate case. Of course growth on NiCr does not result in perfect alignment of the easy axes but leads to a 2D random orientation of the IrMn in the plane due to the cubic structure of the NiCr.

Nonetheless it is clear from the data in Table 1 that texturing the anisotropy into the 2D random case can enhance the effective anisotropy constant by an order of magnitude. Hence the values of $K_{AF}$ quoted are an effective value and, at this time, there is no known mechanism to determine the value of the intrinsic value of $K_{AF}$ unless perfect alignment along a single axis could be achieved. Were that to be possible it would not be unreasonable to envisage that a further increase in anisotropy by perhaps another order of magnitude might result.

### 3.2   Cubic Anisotropy

In addition to the reduction in the energy barrier to reversal associated with cubic anisotropy by a factor 4 it is also the case that there are a multiplicity of easy directions in which the magnetization can lie in the case of a ferromagnet. For example for a single domain particle of a material exhibiting cubic anisotropy and with easy axes along (100) type directions, assuming that the particle is blocked i.e. measurement is made at low temperature, this results in a loop squareness $M_r/M_s$ = 0.83 for K>0 and 0.87 for K<0 [30]. These are calculated values at T=0 but have been partially verified by measurements below 4.2K [31]. Such measurements are for the case where the (100) directions are themselves randomly oriented and therefore should be compared with the value of the squareness of the uniaxial case for random orientation which is $M_r/M_s$=0.5 [16]. These values are indicative of the nature of cubic anisotropy in single domain ferromagnetic grains which presumably would be replicated for the case of AF grains.

There is no obvious way to undertake this comparison directly. However for the case of fcc IrMn$_4$ it is the case that if there are a set of (111) planes in the plane of the substrate then there should be a further set of (111) planes at an angle of 72° to the plane. Of course for the individual grains of IrMn to lie oriented along these other directions it would be necessary for the grains to grow with a 3D random texture i.e. on a Cu seed layer.  Accordingly as part of a separate piece of work, a sample of such a structure was grown where the F layer consisted of a CoPt multilayer in an attempt to induce a perpendicular anisotropy.  The exact structure grown is show in Figure 6 [32].

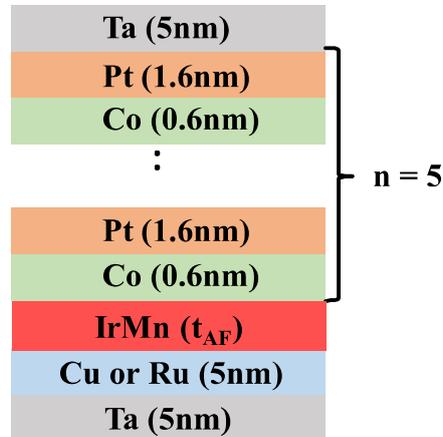

Figure 6. Schematic structure of the Co/Pt multilayer structure used to exchange bias IrMn out-of-plane.

Measurements of the induced exchange bias, always undertaken at a temperature where the system was free of thermal activation ($T_{NA}$), were then made following the setting i.e. the alignment of the AF orientation at different angles to the plane of the film.  The results are shown in Figure 7 and clearly indicate that there is a weak peak in the exchange bias for the sample grown on a Cu seed layer in the vicinity of 72°.  An exact replication of the precise angles in the crystal structure would not be expected because whilst X-ray diffraction indicates that the system exhibits 3D random texture we cannot be certain that there is not some grain size dependence to the random orientation where perhaps larger grains are more likely to be textured in other than a random direction.

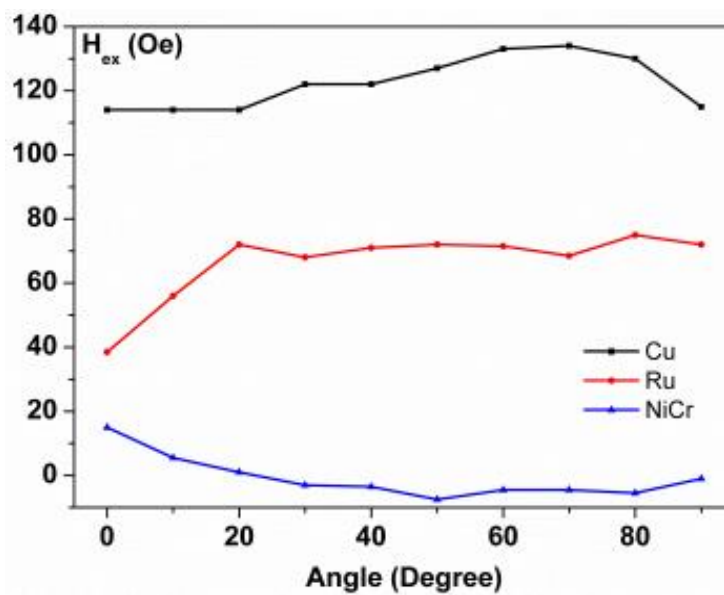

Figure 7. Loop shift as a function of setting angle for sample deposited on different seed layers [32].

This result essentially proves that IrMn$_4$ exhibits cubic anisotropy and therefore should not follow the York Model of Exchange Bias in which uniaxial anisotropy is assumed. Also the energy barrier should be reduced by 75% and so the origin of the high quality fits obtained using the York Model is unclear.

### 3.3    Rotational Effects

One way in which the nature of anisotropy can be determined is by undertaking measurements as a function of angle. In the previous section the angle at which the AF layer was set was probed and indicated the presence of other available (111) planes in an untextured layer of IrMn$_4$. We can now consider the nature of the anisotropy induced by an in-plane setting process where the AF is set at a fixed angle and properties probed as a function of angle relative to that direction [33]. Figure 8(a) shows a set of data for such a system where in this case, the sample was grown without a seed layer i.e. the IrMn will be untextured and was of thickness 5nm.

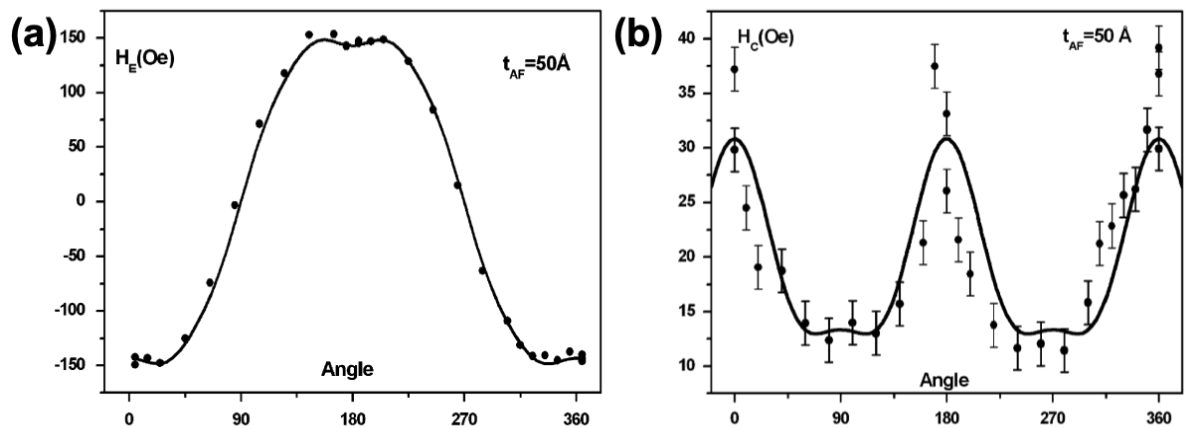

Figure 8. (a) Exchange bias and (b) coercivity as a function of the angle of measurement [33].

The CoFe layer was 10nm thick as the grain size was relatively small in these samples supplied by an external collaborator and hence no grain size distribution could be obtained. The measurements were made at room temperature after the AF was set with the F layer saturated at a temperature of 225°C. Figure 8(a) clearly shows the expected peak in the value of H$_{ex}$ at the setting angle of 180°. Surprisingly the peak is quite broad and would not fit a smooth cos θ function. This is probably due to some irregularity in the setting process but nonetheless the data clearly shows a strong unidirectional anisotropy as there is a significant minimum at angles of π either side of the setting direction. In contrast Figure 8(b) shows the variation of the coercivity of the system measured simultaneously with the loop shift. The behavior is markedly different showing clear uniaxial as opposed to unidirectional anisotropy. This data alone shows that the loop shift H$_{ex}$ and the coercivity H$_c$ have markedly different behavior and therefore must have different origins. We have proposed previously that whilst the exchange bias is controlled by the anisotropy of the bulk of the AF grains, the coercivity must originate from interface spins or spin clusters whose behavior is largely independent of the bulk of the AF grains. This topic will be the subject of a separate work [34]. However when considering anisotropy in exchange bias systems it is important to note these two different anisotropic effects clearly have separate origins.

## 3.4 Anisotropy and Setting

From the previous discussion we have seen that the setting process in which the AF material is field annealed in the presence of the exchange field from the F layer induces a unidirectional texture in the AF grains. However the relationship between the resulting unidirectional anisotropy and the original anisotropy of the AF grains might be different. We have reported two experiments in which again we grew an IrMn layer on top of either a Cu or a NiCr seed layer thereby inducing random 3D texture or random 2D in-plane texture to the (111) planes of the IrMn, respectively. We also grew an F layer consisting of CoFe directly on top of each type of seed layer without the presence of an AF layer. Measurements were then made with the AF layer unset and after setting. The resulting data is shown in Figure 9(a) and 9(b) [35].

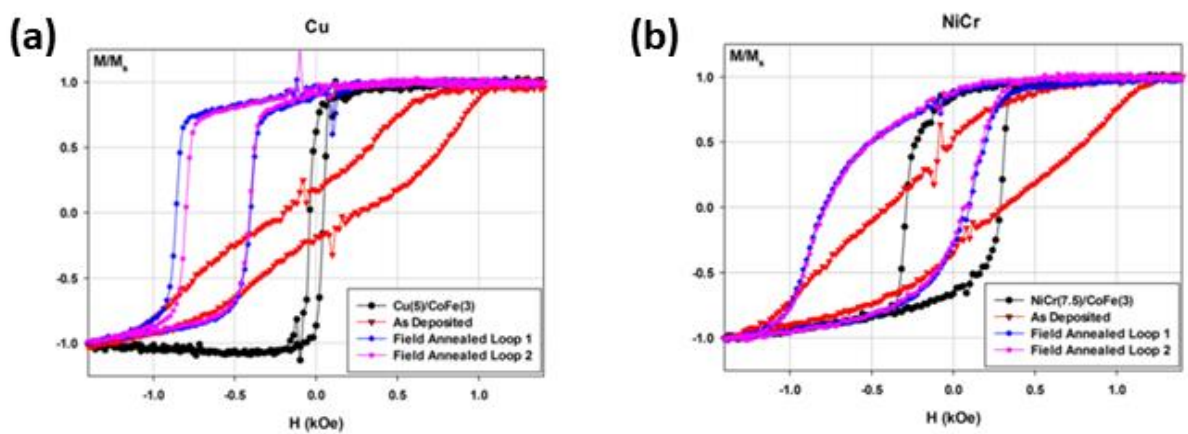

Figure 9. Hysteresis loops for a sample deposited on an (a) Cu and (b) NiCr seed layer [35].

In the case of samples grown on both forms of seed layer the mere presence of the AF layer, albeit not set, beneath the F layer induces a dramatically sheared hysteresis loop. It must be remembered that the data that is seen reflects the behavior of the CoFe layer which ordinarily appears as shown in the graphs (black curves) as a relatively highly exchanged coupled layer where the reversal mechanism will be by nucleation and domain wall movement. The presence of the AF layer then appears to introduce a very wide distribution of domain wall pinning strengths which presumably relate to the anisotropy of the AF grains which would pin the domain walls. This is consistent with the resulting shift in the coercivity of the ferromagnetic layer which increases substantially more in the case of the (111) textured AF grains on NiCr than in the case for those grown on Cu. From these data it is not possible to discern the exact nature of the anisotropy of the AF grains but based on the result of varying the setting angle we know that they behave as if they are cubic with an easy plane along (111).

The effect of setting the AF is then seen to have a quite dramatic effect. Firstly, referring to the sample grown on a Cu seed (Fig 9(a)), the loop becomes highly square and strongly shifted giving an exchange bias of ~700 Oe but also the well-known first loop training phenomenon appears. We have suggested that first loop training is due to the irreversible reorientation of spins or spin clusters at the interface and will be the subject of a separate work [34]. As discussed previously the coercivity of the exchange biased loop is due to interfacial spin effects and hence is not relevant to the discussion of the anisotropy of the bulk. What is clear in this case is that as well as the loop being shifted significantly, the reversal mechanism is initiated by a small amount of domain rotation followed by a nucleation and rapid domain well propagation event with again a small amount of rotation towards negative saturation.

For the sample grown on the NiCr layer (Fig 9(b)) the loop is dramatically different to that for the sample grown on a Cu seed layer. Clearly there is now no training effect. That implies that the increased anisotropy achieved by forcing the IrMn grains to lie with (111) planes in-plane has not only affected the bulk of the AF grains but also the spins at the interface which we believe are responsible for training. However there is now also a very wide distribution of domain wall pinning strengths and the reversal process is far more gradual as the anisotropy energy of the AF grains is much greater than any exchange energy promoting domain wall movement. There is also a very large coercivity which means that in some sense the irreversibility of the interface spins has been increased by the increased anisotropy of the IrMn grains. Ironically this now results in a reduced value of the loop shift. None-the-less it is clear that this texturing has resulted in a significant increase in anisotropy.

4.     The Uniaxial Conundrum

Clearly with regard to the most widely studied structure i.e. IrMn/CoFe there is contradictory data regarding the nature of the anisotropy and this may also be the case for other less widely studies systems. The original York Model of Exchange Bias [5] is based upon the fact that the anisotropy is uniaxial and that the behavior of sputtered thin films with grain sizes of around 10nm follows closely that of a classical Stoner-Wohlfarth fine ferromagnetic particle model. Reference to Figure 6 shows the remarkable quality of the fits to the data based on this relatively simple concept coupled to the fact that there may be unset and unstable grains within the particle volume distribution. The quality of the fits and the number of systems that have been studied over the years clearly indicates that such an exchange bias system must exhibit uniaxial anisotropy. However other studies such as those shown in Figure 7 would indicate that the IrMn is cubic and therefore there is a clear conflict with regard to the nature of the anisotropy in the key material IrMn.

The other factor which remains unexplained with regard to the anisotropy of IrMn is the fact that the effective texture especially when the material is grown on a NiCr seed layer, indicates that the (111) planes lie in the plane of the film. However the axes of these planes will be randomly oriented within that 2-D space. Hence again the fact that the York Model of Exchange Bias fits the experimental data using a single value for a uniaxial anisotropy is again contradictory. Hence there must be other factors at play here influencing or even fundamentally changing the nature of the anisotropy.

## 5. Novel Antiferromagnetic Materials

**5.1 Manganese Nitrade (MnN)**

Over the last few years there have been a number of papers published discussing the properties of the equiatomic compound MnN. The majority of these papers have come from the group of Meinert of Bielefeld University[*] and his collaborators [36,37].

There are a number of compounds consisting of MnN but the equiatomic phase is significant because it exhibits AF order via a superexchange mechanism. Unlike the transition metal oxides, when grown correctly the compound MnN crystallizes in a body centered tetragonal (bct) structure. The degree of elongation in the crystal is relatively modest giving an axial ratio a/c of 1.04. The compound is chemically stable and it has been found to be relatively simple to grow in thin film form by reactive sputtering where a controlled amount of N is mixed in with the normal Ar sputter gas and a Mn metal target is used. The resulting tetragonal crystal lattice and the spin structure is shown in Figure 10(a). Figure 10(b) shows the exchange bias loop for the structure shown in Figure 11(a).

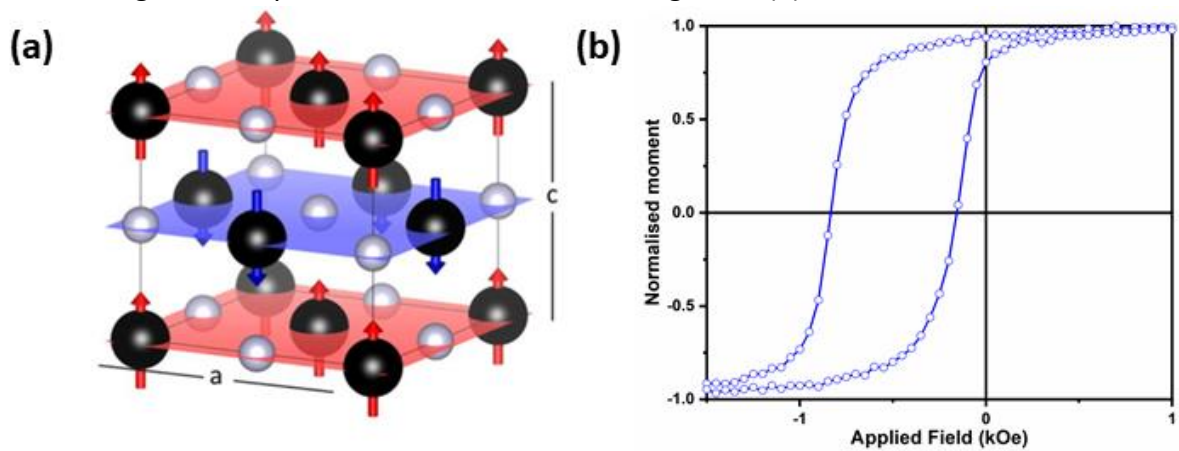

Figure 10. (a) Crystal and spin structure of the equiatomic MnN alloy. Larger spheres with an arrow represent the Mn atoms while the smaller ones denote N (a = 4.256 Å and c = 4.191 Å) and (b) Exchange bias loop for MnN(30nm)/CoFe(1.6nm).

As can be seen from Figure 10(a) the nature of the tetragonal structure of this compound is somewhat anomalous. As shown the length of the a-axis is greater than that of the c-axis as indicated above but then the AF spin order lies along the c-axis rather than along the long axis of the crystal. The origin as to why the shorter c-axis is that along which the spins align is not yet understood. Nonetheless this compound when grown with a layer of CoFe above it is capable of generating a significant degree of loop shift. Figure 11 shows the blocking temperature curve for the sample provided to us by Meinert et al. together with the film structure [38].

---

[*] now at Technical University of Darmstadt.

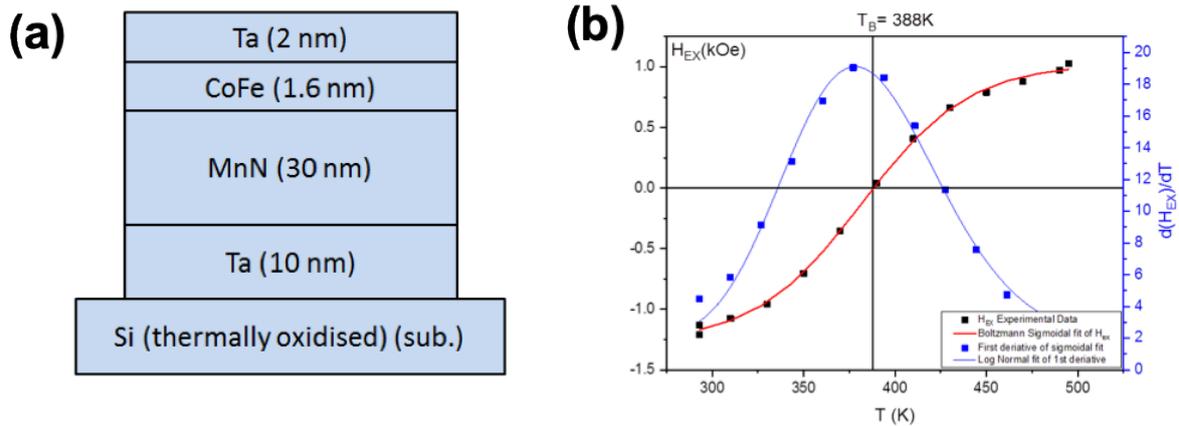

Figure 11. (a) Schematic diagram of the MnN exchange bias structure and (b) The blocking temperatures curve and its derivative for the sample with structure shown in (a).

It is important to note that the data shown in Figure 11(b) is significant in that a relatively large exchange bias of over 1.2 kOe is observed and importantly the shape of the curve in Figure 10(b) is similar to that observed for a highly textured IrMn layer in that the loop is rounded and does not exhibit the properties associated with nucleation and rapid domain wall motion seen in Figure 9(b). Hence it is clear that the MnN grains in this film are imparting a significant increase in anisotropy to the CoFe grains. However it should also be noted that in this case the CoFe layer is very thin at 1.6nm and in order to generate grains having a sufficiently large anisotropy energy barrier (ΔE=KV), it has been necessary to grow a relatively thick (30nm) layer of MnN. A layer of this thickness would preclude the application of this material for example in a read-head device. Nonetheless it is the unusual nature of the anisotropy which is of primary interest here.

Figure 11(b) shows the resulting blocking temperature distribution for the sample shown schematically in Figure 11(a). From this data a median blocking temperature of 388K is observed which again is a value that would be usable in a device albeit with the limitation of the large film thickness. From the data and a knowledge of the grain size, which in this case was obtained from the Scherrer broadening of the X-ray lines, we have made an estimate of the effective anisotropy constant of MnN of $K_{AF}=7\times10^4$ ergs/cc [38]. Whilst this is a modest value compared to that for IrMn which has found to be as large as $K_{AF}=2.7\times10^7$ ergs/cc, nonetheless it is significant that a material with a tetragonal structure has been found to have sufficient anisotropy that, if grown with relatively large grain volumes, a significant exchange bias with the required level of thermal stability can be generated.

**5.2 Heusler Alloys**

In a recent multinational study a search for AF behavior in a range of Heusler Alloys was undertaken in an attempt to find an alternative material to replace IrMn due to the scarcity of Ir [39]. The alloys chosen had the general composition $X_2YZ$ where X and Y are high moment metals and Z is a semiconductor. Many compositions were studied and more than 10 were found to exist in an AF phase some with relatively high values of $T_N$, e.g. $Mn_2VaI$ $T_N > 600K$.

The majority of alloys of this type have a cubic structure with AF order along (100) type directions. Hence other than compositions where X is Mn$_3$ they exhibit cubic anisotropy. In consequence values of T$_B$ are below room temperature in thin film form. For the case of Mn$_3$ alloys the structure is hexagonal and values of T$_B$ above room temperature have been found [39].

Given that there are many thousands of Heusler alloy compositions it is essential that a deeper understanding of the anisotropy in AF materials is developed so that a focused search for new, usable AF alloys can be undertaken.

## 6. Discussion and Conclusion

In this work we have attempted to review the limited information that is known about anisotropy in AF thin film materials. With the rapid development in the field of AF spintronics it is becoming increasingly important for not only new materials to be found or developed but that also some of the underlying mechanisms of the critical parameter of the anisotropy are understood. At present very limited knowledge is available on this subject but it is intended that a new project will be commenced in the near future in our laboratories.

From the data presented, there are clearly a number of relatively well understood effects and significant anomalies observed in the available materials. For example for most AFs which have a cubic structure and are largely transition metal oxides where the exchange interaction is via superexchange, or even alloys where a cubic structure results such as the Heusler alloys which have been studied quite extensively, it is not really surprising that a cubic structure with order along (100) directions gives rise to a relatively low anisotropy and hence materials that, irrespective of the grain size, are not functional at room temperature [39]. However there are clearly other alloys such as FeMn and IrMn where the spin ordering is along (111) planes. This gives rise to a significant effective anisotropy which, from measurements, appears to be of a uniaxial character. However this should not be the case because the (111) directions are in essence a cubic structure. Nonetheless both of these compounds are found to exhibit not only a uniaxial anisotropy of sorts but also of a very significant value of up to 2.7x10$^7$ ergs/cc. A further anomaly occurs in that the optimum exchange bias and hence the optimum anisotropy is found for the off-stoichiometric alloy closer to IrMn$_4$.

In the compound MnN again a worthwhile but not highly significant anisotropy is observed in a tetragonal compound which would be expected because for example PtMn is also tetragonal and is known to have a large anisotropy. However in the case of MnN the easy direction now lies perpendicular to the long axis. Hence it is clear that whilst some understanding of the anisotropy of AFs can be obtained by comparison to the ferromagnetic case there is a clear need for further study to understand the behavior in many AF alloys and compounds.

## 7. Acknowledgments

The authors are happy to acknowledge the long standing and ongoing support and collaboration with Seagate Technology in Northern Ireland. The authors also acknowledge the contribution from Marcus Meinert now of Technical University of Darmstadt who provided the initial samples for our study of MnN. Obviously the work described in this paper relates back to more than a decade and contributions to the original work by former students and researchers in the group. Notably Luis E. Fernandez-Outon Universidade Federal de Minas Gerais, Brazil, is also gratefully acknowledged.

The data that support the findings of this study are available from the corresponding author upon reasonable request.